\documentclass[aps,showkeys,showpacs,twocolumn]{revtex4}
\usepackage{mathrsfs}
\usepackage{booktabs}
\usepackage{amssymb}
\usepackage{amsmath}
\usepackage{bm}
\usepackage{graphicx}
\usepackage{natbib}
\usepackage{rotating}
\usepackage{xcolor}

\begin{document}
\title{Non-Hermitian Twisting Theory under the open boundary condition}

\author{Chen-Hao Zhao$^1$}
\author{Jia-Rui Li$^1$}
\author{Yuping Tian$^1$}
\author{Wei-Jiang Gong$^1$}

\affiliation{1. College of Sciences, Northeastern University, Shenyang 110819, China}
\date{\today}
\begin{abstract}
The non-Hermitian skin effect (NHSE) is a hallmark of non-Hermitian system, yet its generalized Brillouin zone (GBZ) description is restricted to periodic systems. We develop a site-resolved theory via a local scaling transformation (LST), introducing local twisting $T_n$ to quantify metric operator $\xi$ nontriviality. This elucidates the NHSE's origin and uncovers the generalized multiple-channel skin effect (MCSE). Exploiting $T_n$'s translational independence, we define the Zahlen-Brillouin Zone (ZBZ), extending non-Hermitian band theory to nonperiodic and disordered lattices. Furthermore, we unify the $\xi$ with GBZ Riemannian geometry, establishing the metric and state correspondence (MSC) as the principle for real-space localization. With a global skin index $\mathbf\Gamma$ for phase transitions, our results provide a universal paradigm for non-Hermitian physics in both crystalline and amorphous media.

\end{abstract}

\maketitle

\bigskip

\textit{Introduction}-During the past years, non-Hermitian Hamiltonian has been one of research hotspots in the field of quantum physics, including the concept of the $\cal PT$ symmetry system \cite{1,2,3} and exceptional point \cite{Yang2021,Li2026}. The non-Hermitian skin effect (NHSE)\cite{Yang2024,Huang2026,Yang2026}, characterized by the anomalous collapse of bulk eigenstates toward boundaries, represents a fundamental departure from conventional Bloch band theory \cite{4,5,6,7,8}, which has been made in classical wave platforms, including acoustic \cite{Zhang2021a,Gu2022,Zhang2021b,Zhou2023}, optical\cite{Weidemann2020,Xiao2020,Lin2022a,Liu2024,Lin2022b}, mechanical\cite{Ghatak2020,Brandenbourger2019,Chen2021,Wang2022}, and electrical circuit systems\cite{Helbig2020,Hofmann2020,Liu2021,Zou2021,Zhang2023}. These experiments have vividly demonstrated the existence of non-Hermitian skin modes in various systems. To describe this phenomenon, the generalized Brillouin zone (GBZ) framework was established, effectively restoring momentum-space topology by deforming the standard Brillouin zone (BZ) into the complex plane \cite{12,13}. However, the GBZ relies fundamentally on the existence of a complexified wave vector, a concept that is well-defined only in translationally invariant systems\cite{14,15}. This crystalline constraint poses a significant hurdle: the GBZ becomes fundamentally ill-defined in the presence of disorder, quasiperiodicity, or amorphous geometries, where real-space localization remains a pervasive yet poorly understood feature\cite{51,54} . Furthermore, a site-resolved understanding of the physical origin of the NHSE, along with whether it represents the most general form of non-Hermitian localization, remains elusive.

In this Letter, we address these limitations by formalizing a site-resolved theory of the NHSE rooted in the interplay between non-Hermitian symmetry and real-space metricity. By introducing the Local Scaling Transformation (LST), we define a fundamental physical quantity, the local lattice twisting $T_n$, which quantifies the site-specific nontriviality of the non-Hermitian metric operator $\xi$ which will retreat to pseudo-Hermitian symmetry. This framework not only elucidates the real-space origin of the NHSE but also unveils a more generalized phenomenon: the multiple-channel skin effect (MCSE). In the MCSE, the competition or cooperation between local twisting channels facilitates complex localization patterns, such as internal or bipolar localization, that transcend the conventional boundary-constrained skin effect.

Crucially, the translational independence of $T_n$ allows us to extend non-Hermitian band theory beyond the periodic limit. We define the Zahlen-Brillouin Zone (ZBZ), a complexified momentum manifold that remains valid for arbitrary nonperiodic or disordered lattices. By analyzing the ZBZ, we reveal a deep geometric unification between the $\xi$ metric and the Riemannian metric of the GBZ curve. This leads us to establish the metric and state correspondence (MSC), which identifies the real-space inner-product structure as the fundamental principle underlying the efficacy of complex momentum-space descriptions.

Finally, we introduce a global skin index $\Gamma$, derived from the local twisting $T_n$, as a robust diagnostic tool for quantifying non-Hermitian nontriviality and benchmarking phase transitions. Our work provides a universal paradigm for characterizing non-Hermitian physics across the spectrum from crystalline order to amorphous disorder, bridging the gap between spectral topology and real-space manifestation.

\begin{figure}
\begin{center}\scalebox{0.1}{\includegraphics{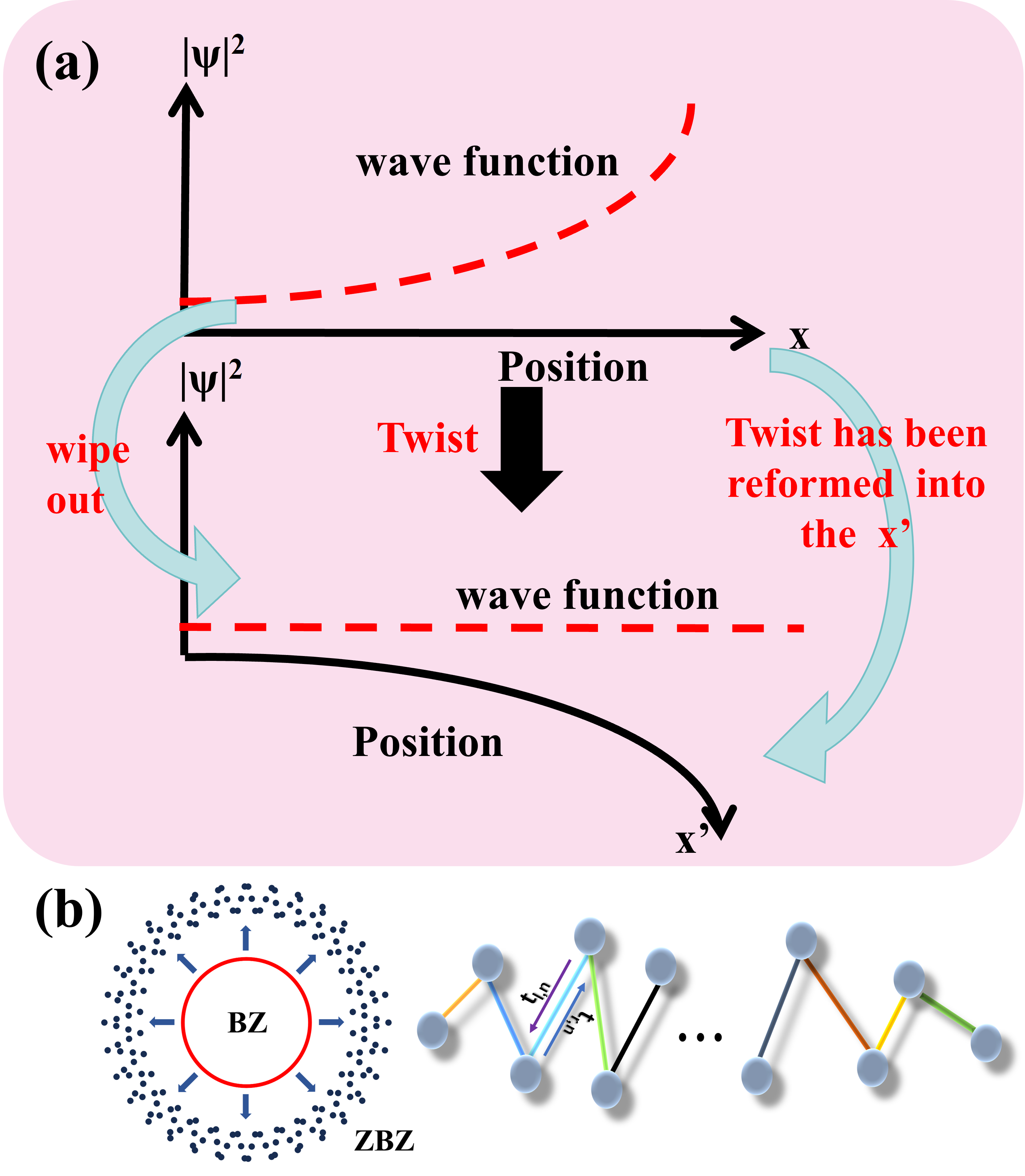}}
    \caption{Schematic of lattice twisting and the Zahlen Brillouin zone (ZBZ). (a) The metric of the non-reciprocal Hamiltonian is modified (see text) to absorb the exponential growth of the wavefunction into the coordinate, removing the accumulated twisting and restoring a Bloch-wave form via a LST. (b) The Brillouin zone is discretized to incorporate nonreciprocity, defining the ZBZ.}
    \label{fig:placeholder}
\end{center}
\end{figure}

\textit{Nonreciprocal is a twisting}-Nonreciprocal systems with periodic structures, such as the Hatano–Nelson (H-N) chain and the nonreciprocal Su–Schrieffer–Heeger (SSH) model, can be understood within the framework of non-Bloch band theory. In these cases, non-Hermitian topological conditions can be derived via similarity transformations or the GBZ, and the NHSE is properly captured. However, such approaches apply only to periodic lattices, leaving the more general nonperiodic or disordered systems beyond reach. As a demonstration, we consider the following generic nonreciprocal Hamiltonian without assuming any translational invariance:
\begin{align}
     H&=\sum [t_{l,n}c^{\dagger}_{n}c_{n+1}+t_{r,n}c^{\dagger}_{n+1}c_{n}].
\end{align}
where $c^{\dagger}_{n}$ is the creation operator for $n$-th unit cell, $t_{l,n}$ and $t_{r,n}$ are the asymmetric intracell hoppings.

To absorb the nonreciprocity of the system as illustrated on Fig.1 (a), we introduce a new set of basis states $d_n^{\sharp}/d_n$ $H(c_n^{\dagger}/c_n)\rightarrow H'(d_n^{\sharp}/d_n)$, which has
\begin{align}
    d_n^{\sharp}d_{n+1}=T_n^{-1}c_n^{\dagger}c_{n+1},\\\notag
    d_{n+1}^{\sharp}d_{n}=T_nc_{n+1}^{\dagger}c_{n}.
\end{align}
Here $T_n$ represents the nonreciprocity in each cell, which refer to the twisting.
The relation between the new and original bases reads
\begin{align}
    d_n=Oc_nO^{-1},\,\,\,\,d^{\sharp}_n=Oc^{\dagger}_nO^{-1}.
\end{align}

However, the new basis does not necessarily satisfy Hermitian conjugation. We thus assume a generalized conjugation defined as $d_n^{\sharp}=\xi^{-1} d_n^{\dagger}\xi$, $\xi$ encodes the full nonreciprocal information, capturing the site-resolved nonreciprocity and defining a twisting $\xi$-metric. In the trivial case, $\xi=I$, the above relation reduces to $d^{\sharp}=d^{\dagger}$, recovering the conventional Hermitian Hamiltonian; In the nontrivial case with $\xi\neq I$, the structure of the Hilbert space is non-trivially modified, which captures the intrinsic nature of nonreciprocal non-Hermitian systems.

For a tridiagonal matrices, the Eq.~(2) construction requires a site-dependent local scaling transformation (LST) $O$, which takes the form
\begin{align}
     O=e^{-\sum_{n=1}^{N-1}\mu_n c_n^{\dagger}c_n}.
\end{align}
Here, $\mu_n$ denotes a local scaling factor, representing the accumulated nonreciprocity of the Hamiltonian space at site $n$. Without loss of generality, we set $\mu_0 = 1$. Figs.2 (a-b) illustrates the matrix elements $O_n$ of $O$ (red circle) alongside the wavefunction profile $\sum_n \psi_{n,j}/N$ (cyan line) for comparison. Figs.2 present specific wave functions, confirming the action of $O$, including localization, directionality and form.

\textit{Twising for wavefunction}-Substituting Eq.~(4) into Eq.~(3), one can derive the representation of twisting
\begin{align}
    T_n\equiv\text{exp}(\mu_{n+1}-\mu_{n}).
\end{align}
This indicates that $T_n$ captures the local nonreciprocity at each lattice site.
Combining Eq.~(2) with the Hamiltonian Eq.~(1), the exact relation between the new and original bases is readily obtained [SM]:
\begin{align}
    d_n=\sqrt{\xi_n}\cdot c_n=\prod_{q=1}^{n-1}T_q c_n.
\end{align}
Here $T_n=\sqrt{\frac{t_{r,q}}{t_{l,q}}}$ $\xi_n=|O_n|^2$.
The global twisting effect of the entire lattice is uniquely determined by the collection array of all local twistings:
\begin{align}
    \textbf{T}=\{1,T_1,T_2,...,T_{N-1}\}.
\end{align}
$\textbf{T}$ fully capturing the spatial profile of local nonreciprocity in nonperiodic lattices.
The local twisting $T_n$ directly quantifies the nonreciprocal asymmetry between the
$n$-th and $(n+1)$-th sites. $T_n$=1 corresponds to a Hermitian site with $t_{l,n}=t_{r,n}$ and no twisting, whereas $T_n\neq1$ indicates a nonreciprocal site with lattice twisting.

From Eq.~(6), $|\bar{\psi}\rangle = \sum_{n} \bar{\phi}_n |d_n\rangle$, with $|d_n\rangle = (\prod_{q=1}^{n-1} T_q) |c_n\rangle$, formalizes the mapping from the original non-Hermitian Hamiltonian $H$ to its Hermitian counterpart $\bar{H}$. Here, the local twisting $T_n$ serves as a site-resolved measure of nonreciprocity, where $T_n=1$ denotes the Hermitian limit and $T_n \neq 1$ signifies a local amplification channel: $T_n > 1$ ($T_n < 1$) rescales the wavefunction amplitude rightward (leftward). In this framework, the original wavefunction amplitude $\phi_n$ is rigorously linked to the uniform Bloch amplitude $\bar{\phi}_n$ of the Hermitian system via
\begin{equation}
    \phi_n = \bar{\phi}_n \cdot \sqrt{\xi_n} = \bar{\phi}_n \cdot \prod_{q=1}^{n-1} T_q.
\end{equation}
Equation (8) reveals that the localization envelope of the NHSE is entirely encoded in the integrated effect of local twisting.

We categorize the resulting phenomena into two regimes based on the spatial distribution of $T_n$. Under \textit{unidirectional twisting} (UT), where $T_n \gtrless 1$ uniformly across the lattice, the cumulative product drives a monotonic exponential evolution, resulting in the standard boundary NHSE [Fig.~3(a)(iv), Figs.~3(b)(ii-iii)]. Conversely, \textit{competing twisting} (CT) involves the spatial coexistence of $T_n > 1$ and $T_n < 1$. This regime triggers an interplay between opposing twisting channels, facilitating the \textit{multiple-channel skin effect} (MCSE). Within the MCSE framework, the competition or cooperation among sites enables localization at arbitrary positions [Fig.~3(a)(ii)], bipolar-NHSE [Fig.~3(a)(iii)], or even critical states [Fig.~3(a)(i), Fig.~3(b)(i)]. Notably, the MCSE [Fig.~3(a)(ii)] transcends the conventional domain-wall picture; while domain-wall states exhibit directional localization with asymmetric decay, the MCSE is characterized by balanced local amplification and attenuation across the lattice without a preferred global direction.

\begin{figure}
\begin{center}\scalebox{0.08}{\includegraphics{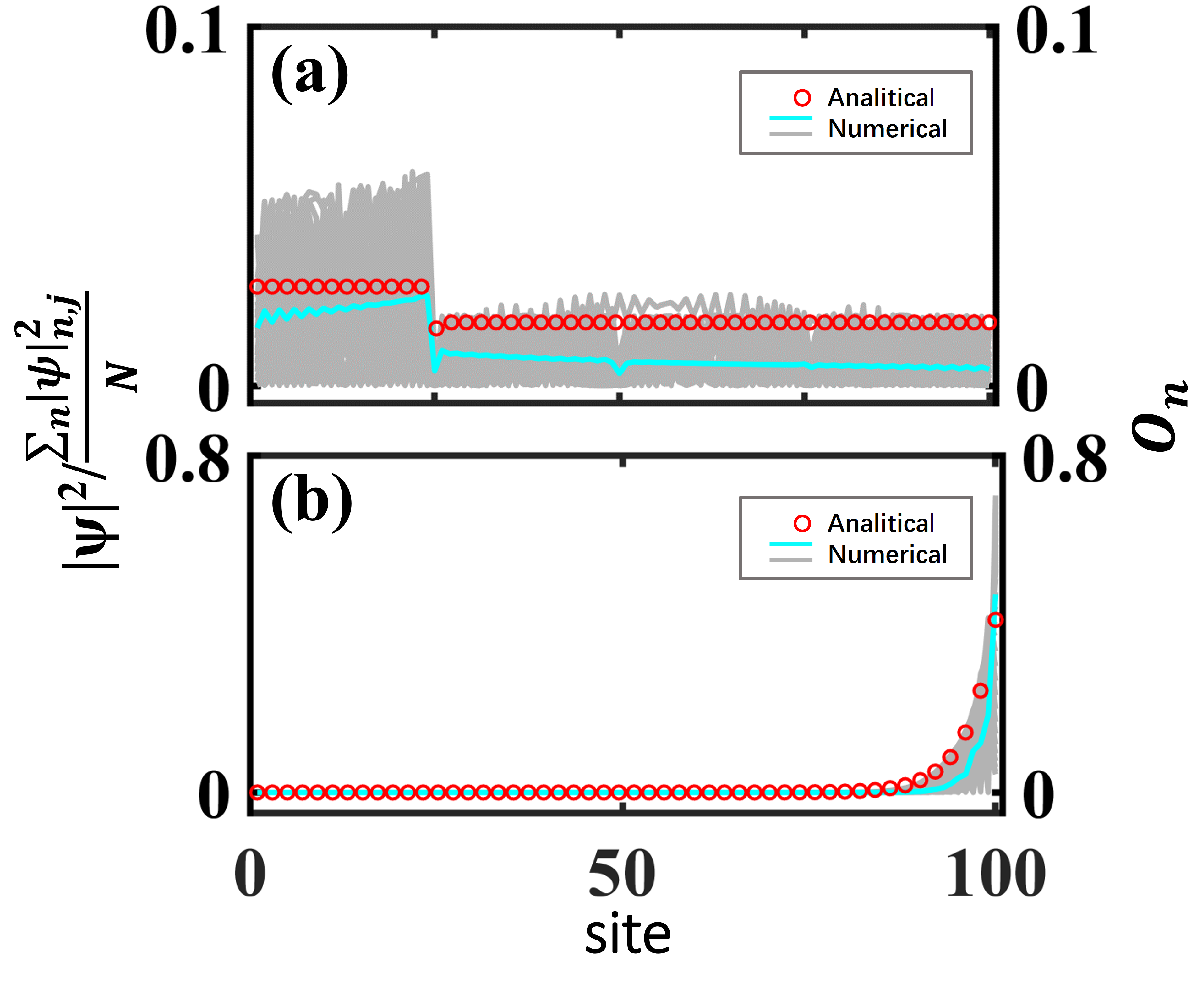}}
    \caption{Schematic of twisting accumulation. The local scaling transformation $O_n$ (red circle) captures the cumulative effect of lattice twisting, which underlies the NHSE. Two representative cases are shown: (a) $H_1=t_1(c^{\dagger}_{A,n}c_{B,n}+h.c.)+t_2(c^{\dagger}_{B,n}c_{A,n+1}+h.c.)+\gamma c_{A,13}^{\dagger}c_{B,13}-\gamma c_{B,13}^{\dagger}c_{A,13},\, t_{r,25}\rightarrow (t_1-\gamma), t_{l,25}\rightarrow (t_1+\gamma), t_{r,n}=t_{l,n}=t_1$, with a single nontrivial twisting $T_{25}$, and (b) $H_2=\sum_{n=1}^{N-1}[t_1(c_{A,n}^{\dagger}c_{B,n}+h.c.)+t_2(c_{B,n}^{\dagger}c_{A,n+1}+h.c.)]-\sum_{n=1}^{N/2}\gamma c^{\dagger}_{n}c_{n+1}+\sum_{n=N/2+1}^{N}\gamma c_{n+1}^{\dagger}c_{n}$, with region-dependent global twisting. The grey line and cyan line are the wavefunction densities and  wavefunction density profiles for $H_1$ and $H_2$. Parameters N=100, $t_2$=1, $t_1=0.9$; $H_1$: $\gamma=2$, $H_2$: $\gamma=1.2$.}
    \label{fig:placeholder}
\end{center}
\end{figure}

\begin{figure*}
\begin{center}\scalebox{0.082}{\includegraphics{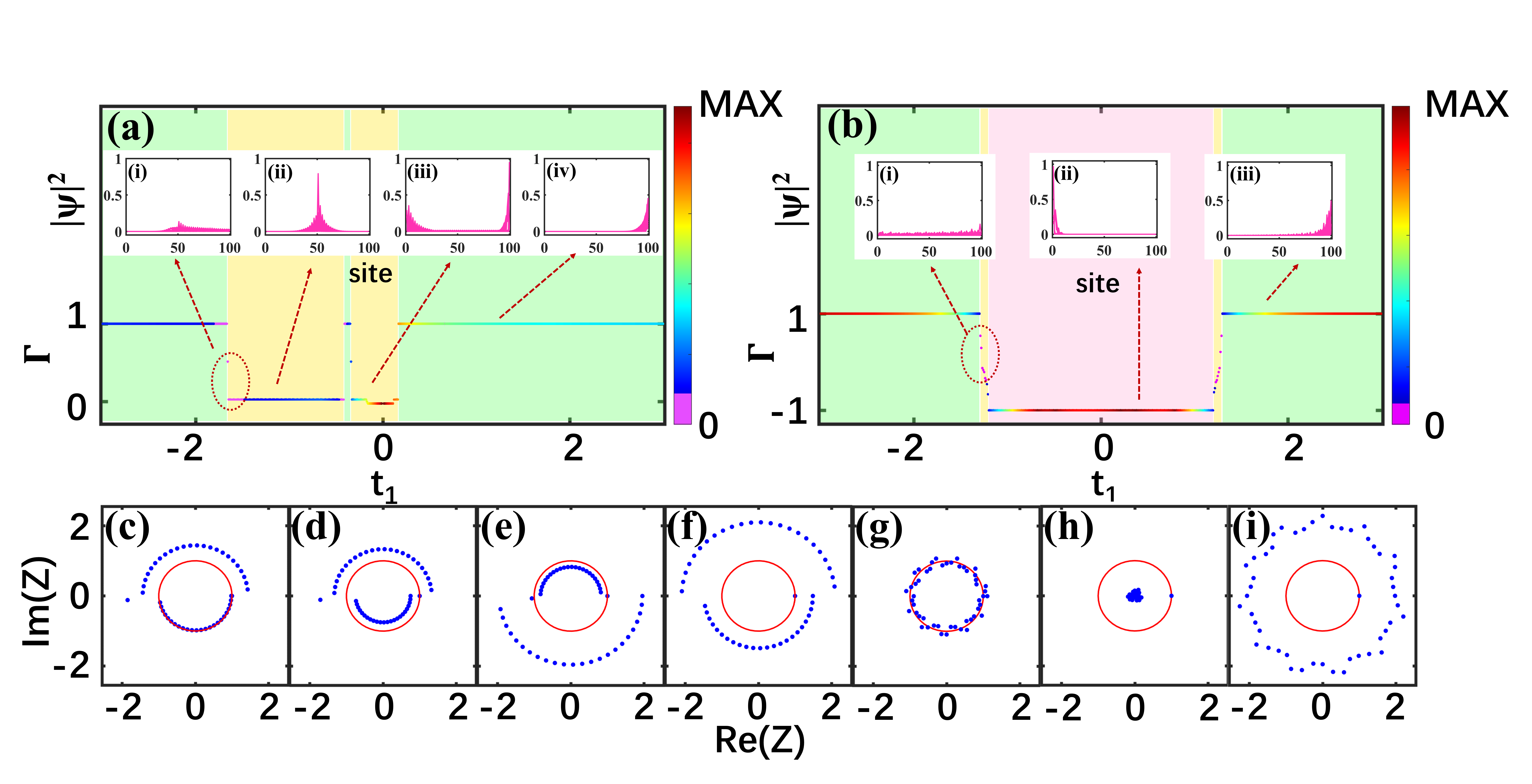}}
    \caption{Multiple-channel skin effect (MCSE) in non-periodic system $H_2$ and quasi-periodic system $H_3$ ($H_3=\sum_n t_1c_{A,n}^{\dagger}c_{B,n}+\sum_n t_2(c_{B,n}^{\dagger}c_{A,n+1})+\sum_n [t+\delta\text{cos}(2\pi bn+\phi)](c_{B,n}^{\dagger}c_{A,n}+c_{A,n+1}^{\dagger}c_{B,n}).$) (a) The skin index $\mathbf{\Gamma}$ for $H_2$. Representative configurations: (i-iii) correspond to typical MCSE, with (iii) also referred to as the bipolar NHSE, while (iv) represents the conventional NHSE, with the corresponding wavefunction localization (color-coded by the right color bar). The parameters are $t_1=-1.57,-1,-0.2,2$, respectively. (b) The skin index $\mathbf\Gamma$ for $H_3$, showing (i) correspond to MCSE, (ii-iii) represent the conventional NHSE, together with the corresponding wavefunction localization (color-coded by the right color bar). $\mathbf\Gamma$ is divided two different phase, Antagonistic phase (AP, yellow region for $\Gamma\in(-1,1)$) and completed phase (CP, green for $\Gamma=1$, pink for $\Gamma=-1$). Remarkably, the purple dot is a critical phase, means most of states turns to the expended-like states in AP (AP-E), which is indicted by $\{\bar O_n\}_{\text{max}} \sim \mathcal O(N)^{-1}$. (c-i) Corresponding ZBZ (blue dots) for the cases in (a)(i-iv) and (b)(i-iii). In the AP phase, the ZBZ distributes both inside and outside the BZ (red line), whereas in the CP phase it lies entirely on one side of the BZ.}
    \label{fig:placeholder}
\end{center}
\end{figure*}

\textit{Zahlen-Brillouin zone}-Having established the real-space correspondence between lattice twisting and the NHSE, we now extend the framework by introducing the Zahlen Brillouin zone (ZBZ). The conventional GBZ theory relies on translational invariance and is therefore restricted to periodic systems. By contrast, the ZBZ is constructed in a discrete, local $z_n$-space that does not assume translational symmetry, while remaining compatible with the GBZ framework. This formulation removes the reliance on translational invariance and provides a unified description for periodic, nonperiodic, and disordered lattices.

Exploiting the Bloch wave $\bar{\psi}_n$ with the translational symmetry and Eq.~(8), we can analyze the phase accumulation of the wavefunction under OBC. The spatial translation of the wavefunction at the $n$-th lattice site consists of two contributions: the amplitude rescaling induced by the $T_n$, and the local Bloch phase $e^{iz_n}$. Here $z_n = 2n\pi/N \in [0,2\pi]$ denotes the discrete local momentum associated with the $n$-th site, covering the entire BZ range. Consequently, for the wavefunction $\psi(r)$, the amplitude at site $r+j$ is strictly related to that at site $r$ by [sm]
\begin{align}
\psi(r+R_j)&=\prod_{q=1}^{j}[T_qe^{i\frac{n2\pi}{N}}]\psi(r).
\end{align}

In the thermodynamic limit $N\rightarrow\infty$, the Zahlen Brillouin zone (ZBZ) is defined as the discrete complex set
\begin{align}
    Z=\{Z_n\}=\{T_ne^{iz_n}|n=1,2,...N-1\}
\end{align}
Here $|Z|=T_n$ gives the magnitude of the ZBZ element, characterizing the local lattice twisting at site $n$, while $\text{arg}(Z_n)=z_n$ specifies the discrete local momentum. For a periodic nonreciprocal system, translational invariance requires the local twisting to be identical at all sites, $T_n=T=\beta$, independent of $n$. In the thermodynamic limit, the discrete local momentum $z_n$ becomes the continuous Bloch momentum $k\in[0,2\pi]$ in which case the ZBZ reduces to: $ Z=\{Te^{ik}|k\in[0,2\pi]\}$, this shows that the ZBZ is a natural generalization of the GBZ and reduces exactly to the conventional GBZ in periodic systems. Therefore, the ZBZ reproduces the GBZ framework of the NHSE in periodic nonreciprocal systems. The MCSE arises from the CT, for which the ZBZ spans the Brillouin zone [Figs.3(c-e),(g)]. In contrast, when UT emerges, corresponding to the NHSE, the ZBZ expands beyond (contracts within) the BZ [Fig.3(f), (h-i)]. These behaviors correspond to the MCSE and NHSE shown in Fig.3(a)(i-iv) and Fig.3(b)(i-iii). In the limit $T_n=1$, the ZBZ reduces to the standard BZ, thereby elucidating how momentum-space quantities can fundamentally capture real-space states localization phenomena.

\textit{Metric and states correspondence}-The compatibility between the ZBZ and GBZ in periodic systems reveals an intrinsic physical unification, wherein the inverse metric $\xi$ maps rigorously onto the Riemannian metric of the GBZ/BZ manifold. This correspondence identifies the real-space inner-product structure as a direct manifestation of the complex momentum-space geometry. For a periodic non-reciprocal lattice, the GBZ forms a circle in the complex plane with radius $r =\sqrt{ t_r/t_l }= T$, whose Riemannian metric tensor is $g_{kk} = |d\beta/dk|^2 = r^2 = T^{2}$, the diagonal elements of the metric satisfy
\begin{align}
    \xi_n = g_{kk}^{n}.
\end{align}
So that their scaling properties are entirely determined by the nonreciprocal asymmetry. We further demonstrate that $\xi$ serves as the natural metric for the orthonormalization of generalized Bloch states on the GBZ [SM], with strict orthonormality being recovered only under this weighted inner product. This correspondence elucidates the efficacy of the GBZ in characterizing the NHSE, as its momentum-space metric intrinsically encodes the real-space operator $\xi$ that fundamentally dictates the eigenstate localization profile. This integration of real-space metricity into a complexified momentum manifold restores the predictive power of band theory under nonreciprocity. This fundamental link between spectral topology and spatial localization formalizes what we term the metric-state correspondence (MSC).


\textit{Skin index}-Based on the directional distribution of the local twisting $T_n$, we introduce a directly computable criterion to diagnose the NHSE type (left, right, or multiple-channel) and its localization strength in arbitrary systems. Since this quantity depends solely on $T_n$ and does not assume periodicity, it provides a universal diagnostic. We therefore define the normalized fundamental skin index as
\begin{align}
    \Gamma=\sum_{n=1}^{N-1} \frac{sign[ln(T_n)]}{N-1}
\end{align}
Physically, $\Gamma$ represents the average twisting direction of the lattice. It is strictly bounded in the interval $[-1,1]$, quantifying the dominant direction of the global twisting.
To characterize both the NHSE type and its localization strength, we introduce the maximal normalized cumulative twisting weight $\{\bar O_{n}\}_{\text{max}}$, which takes values in the range $(0,1]$, with larger values indicating stronger localization and define a two-dimensional extended skin index as:
\begin{align}
    \mathbf{\Gamma}=\begin{bmatrix}
        \Gamma\\
        \{\bar O_{n}\}_{\text{max}}
    \end{bmatrix}.
\end{align}
where the maximal cumulative twisting weight is defined by $\{\bar O_{n}\}_{max}=\{\prod_nT_n/\sum_n\prod_nT_n\}_{max}$. The extended skin index $\Gamma$ establishes a one-to-one correspondence with the NHSE phase, providing a complete classification of the MCSE.

\begin{table}[t]
\centering
\scriptsize
\renewcommand{\arraystretch}{1.6}
\setlength{\tabcolsep}{4pt}
\begin{tabular}{p{1.0cm}| p{2.8cm} p{3.15cm}}
\hline
\hline
\multicolumn{1}{c|}{Attribute} &
\multicolumn{1}{c}{Complete phase (CP)} &
\multicolumn{1}{c}{Antagonistic phase (AP)} \\
\hline
\hline

\centering Skin index
&  $\Gamma=\pm1$ $\{\bar O_n\}_{\max}\to1$
&  $\Gamma\in(-1,1)$  $\{\bar O_n\}_{\max}<1$ \\

\centering Local twisting
&  UT ($\text{ln}T_n>(<)1$)
&  CT (Coexistence of $\text{ln}T_n>1$ and $\text{ln}T_n<1$)\\

\centering NHSE feature
& {\centering Right (Left) NHSE}
& {\centering MCSE;$\{\bar O_n\}_{\max}\sim\mathcal O(N)^{-1}$ AP-E} \\
\hline
\end{tabular}
\caption{Transposed phase classification of the NHSE based on $\Gamma$ and $\{\bar O_n\}_{\max}$. CP includes right (RCP) and left (LCP) NHSE.}
\label{tab:nhse_phase_transposed}
\end{table}

Furthermore, $\Gamma$ facilitates the characterization of the $\xi$ metric's nontriviality. The magnitude $|\Gamma|$ reflects the monotonicity of the metric's diagonal elements, where $|\Gamma| \to 1$ signifies enhanced non-reciprocity and a more pronounced NHSE. Crucially, $\Gamma$ bypasses the requirement of translational symmetry; for arbitrary disordered lattices, $\Gamma$ facilitates a rapid diagnosis of both the type and strength of the NHSE once the local twisting $T_n$ is determined [Figs.3 (a-b)], where the phase transition conditions are min\{$\text{ln}T_n\}<0$ and max $\{\text{ln}T_n\}>0$. Specifically, in Fig.3(a) the AP emerges for $t_1 \in [-1.66, -0.41] \cup [-0.34, 0.17]$, while the CP prevails otherwise. In Fig.3 (b) system turns in AP when $t_1\in[-1.32,-1.20]\cup[1.20,1.32]$. Correspondingly, Table~I summarizes the phase values in units of $\Gamma$. Furthermore, $\Gamma$ is experimentally accessible via measurements of wavefunction localization, density of states, or momentum-resolved spectra, providing a robust quantitative criterion for experimental verification.

\textit{Conclusion}-In summary, we have developed a theoretical framework for non-Hermitian systems by exploiting the Local Scaling Transformation (LST). We introduced the local lattice twisting $T_n$ to quantify the site-resolved nontriviality of the metric operator $\xi$, providing a novel perspective on the physical origin of the NHSE and uncovering a more generalized phenomenon, i.e., the multiple-channel skin effect (MCSE). By leveraging the translational independence of $T_n$, we extended the conventional GBZ theory to the Zahlen-Brillouin Zone (ZBZ), effectively transcending the constraints of periodicity to characterize real-space localization in nonperiodic media and firstly providing the site-resolved interpretation of how nonreciprocity reshapes the BZ. Furthermore, we unified the $\xi$-metric with the Riemannian geometry of the GBZ, establishing the metric-state correspondence (MSC) as the fundamental principle underlying these real-space phenomena. Finally, we introduced the skin index $\Gamma$ as a robust tool for benchmarking NHSE phase transitions. Our work provides a universal platform for understanding non-Hermitian topology across ordered and disordered systems alike.

\end{document}